\documentclass[12pt]{article}

\usepackage{amssymb}
\usepackage{color}
\usepackage{epsfig,amssymb,amsfonts,amsmath,graphicx,dsfont,cite,xfrac}
\usepackage{authblk}
\usepackage{subcaption}
\definecolor{mygray}{gray}{0.5}
\usepackage{multirow}

\usepackage{cite}
\usepackage[colorlinks=true,linkcolor=blue,citecolor=red]{hyperref}

\newcommand{\be}{\begin{equation}}
\newcommand{\ee}{\end{equation}}
\newcommand{\bea}{\begin{eqnarray}}
\newcommand{\eea}{\end{eqnarray}}

\title{Interlace properties for the real and imaginary parts of the wave functions of complex-valued potentials with real spectrum}

\author[${}$]{Alfonso Jaimes-N\'ajera}
\author[${}$]{Oscar Rosas-Ortiz}

\affil[${}$]{\footnotesize Physics Department, Cinvestav, AP 14-740, 07000
M\'exico DF, Mexico}

\date{}
\begin{document}

\maketitle

\begin{abstract}
Some general properties of the wave functions of complex-valued potentials with real spectrum are studied. The main results are presented in a series of lemmas, corollaries and theorems that are satisfied by the zeros of the real and imaginary parts of the wave functions on the real line. In particular, it is shown that such zeros interlace so that the corresponding probability densities $\rho(x)$ are never null. We find that the profile of the imaginary part $V_I(x)$ of a given complex-valued potential determines the number and distribution of the maxima and minima of the related probability densities. Our conjecture is that $V_I(x)$ must be continuous in $\mathbb R$, and that its integral over all the real line must be equal to zero in order to get control on the distribution of the maxima and minima of $\rho(x)$. The applicability of these results is shown by solving the eigenvalue equation of different complex potentials, these last being either ${\cal PT}$-symmetric or not invariant under the ${\cal PT}$-transformation. 
\end{abstract}


\section{Introduction}

It is well known that the spectral properties of the one-dimensional Schr\"odinger operator 
\be
H \psi = -\psi'' + V(x) \psi,
\label{intro1}
\ee
where $V(x)$ is a real-valued, measurable and locally bounded function of $x\in \mathbb R$, can be studied in terms of the Sturm oscillation theorem \cite{Ber91}. It follows from the Sears theorem that this operator is essentially self-adjoint in $C^{\infty}(\mathbb R)$ if $V(x) \rightarrow +\infty$ as $\vert x \vert \rightarrow + \infty$ (the proof and further details can be consulted in Ch.~2 of  \cite{Ber91}). As the zeros of successive eigenfunctions of self-adjoint Schr\"odinger operators interlace \cite{Ber91}, the Sturm-Liouville theory ensures that the related eigenfunctions are complete (see, e.g. \cite{Inc56,Amr05} and references quoted therein).

If the potential $V(x)$ in (\ref{intro1}) is complex-valued then their eigenfunctions are also complex-valued, even if the corresponding eigenvalues are real. Moreover, the eigenfunctions are not complete in the conventional sense anymore, so that the notion of bi-orthogonality \cite{Bre91} is necessary. For potentials that are invariant under space-time reflection \cite{Ben99,Ben05}, the eigenvalue problem has been regarded as the analytic extension of a Sturm-Liouville problem into the complex plane to get heuristic evidence that the eigenfunctions might be complete \cite{Ben00}. However, a rigorous proof of such a property is still absent because it is not clear what space must be used to define completeness in this case \cite{Ben00}. This last would be overpassed by considering that a ${\cal PT}$-symmetric eigenvalue problem can be described using a Hamiltonian that is Hermitian with respect to a given positive-definite inner product \cite{Mos03}. In this form, there must exist a unitary transformation \cite{Mos03} mapping the space of states of the ${\cal PT}$-symmetric Hamiltonian into a new vector space with the appropriate inner product.

For complex-valued potentials with real spectrum that are not ${\cal PT}$-symmetric the interlacing of the zeros of their eigenfunctions has not been studied. The diversity of such a class of potentials is very wide. For instance, it is enough to consider the classification of ${\cal PT}$-symmetric potentials presented in \cite{Lev00}, and to impose the conditions to broke such a symmetry. Another important branch of complex potentials that are not invariant under the ${\cal PT}$-transformation, but have real spectrum, arises from the supersymmetric formulation of Quantum Mechanics \cite{And85,And93,Bag00,Coo01,Mie04,Nic14}, see for example \cite{Can98,And99,Bag01,Ros15}.

Remarkably, the wave functions $\psi(x)$ belonging to real eigenvalues and complex-valued potentials are free of nodes. That is, such functions do not have zeros on the real line. This last means that the real $\psi_R(x)$ and imaginary $\psi_I(x)$ parts of a given wave function $\psi(x)$ do not share any zero in $\mathbb R$. However, they have a series of zeros, individually, when $x$ covers the real numbers. Accordingly, the corresponding probability densities $\rho(x) = \vert \psi(x) \vert^2$ exhibit a number of maxima and minima. Then, to give a description of the behaviour of a particle with one degree of freedom that is subjected to the action of a complex-valued potential, it is necessary to investigate the interlace properties of the zeros of $\psi_R(x)$ and $\psi_I(x)$ in $\mathbb R$. The affirmation holds for any complex potential with real spectrum, no matter if this is either invariant or not invariant under space-time reflection.

In this work, we study some properties that are common in the wave functions of a wide class of complex-valued potentials with real spectrum. In particular, we show that the absence of zeros in the probability densities of these systems is due to the interlacing of the zeros of $\psi_R(x)$ and $\psi_I(x)$ in $\mathbb R$. Moreover, we shall see that the distribution of the maxima and minima of $\rho(x)$ is regulated by such an interlacing. We include examples addressed to show that, in general, for complex-valued potentials which have been generated in arbitrary form, there is not control on the number and the distribution of the zeros of $\psi_R(x)$ and $\psi_I(x)$ in $\mathbb R$. Then we show that this is not the case for the complex potentials that are generated by the supersymmetric approach introduced in \cite{Ros15} because, in such case, the number of zeros is finite and determined by the energy level of the bound state under study. We find that the profile of the imaginary part $V_I(x)$ of these potentials plays a relevant role to get control on both, the zeros of $\psi_R(x)$ and $\psi_I(x)$ in $\mathbb R$, and the distribution of maxima and minima of $\rho(x)$. Namely, $V_I(x)$ is continuous in $\mathbb R$ and its integral over all the real line is equal to zero. The latter is referred to as the condition of {\em zero total area}. Our conjecture is that this profile is universal in the complex-valued potentials with real spectrum which allow such a control on $\psi(x)$ and $\rho(x)$.

As the wave functions $\psi(x)$ are complex-valued, to analyze the interlacing of their zeros it is necessary the analytic continuation of the eigenvalue problem to the complex plane. Some insights have been obtained in \cite{Ben00} for the ${\cal PT}$-symmetric potentials. The conjecture indicated above permits visualize that similar results should be obtained for other complex-valued potentials with real spectrum. The verification of this last affirmation is out of the scope of the present work and will be discussed elsewhere.

The paper is organized in two parts. The first one includes the analysis of complex potentials that are constructed in arbitrary form, this corresponds to the Section~\ref{Inter}. The second part is contained in Section~\ref{darbouxsec} and deals with the complex potentials that are generated by the supersymmetric approach. In each case, the main results are firstly presented and then some applications are given. The examples include ${\cal PT}$-symmetric potentials as well as potentials that are not invariant under the ${\cal PT}$ transformation. For the sake of clarity, the proofs of all the formal results included in Sections~\ref{Inter} and \ref{darbouxsec} are presented in Section~\ref{proofs}. Some final remarks and conclusions are given in Section~\ref{concluye}.

\section{Interlacing theorem}
\label{Inter}

Let us consider the one-dimensional Hamiltonian (in suitable units)
\be
H = -\frac{d^2}{dx^2} + V(x).
\label{H}
\ee
We assume that the potential $V(x)$ is a complex-valued function $V: \mathbb R \rightarrow \mathbb C$ such that the eigenvalue equation
\be
H \psi(x) = E \psi (x)
\label{schro1}
\ee
admits normalizable solutions
\be
\int_{\mathbb R} \vert \psi (x) \vert^2 dx < + \infty
\label{squareint}
\ee
for a given set of real eigenvalues 
\be
E_0 <E_1 < E_2 < \cdots.
\label{es}
\ee
Hereafter we shall use $f_R(x)$ and $f_I(x)$ for the real and imaginary parts of any complex function $f: \mathbb R \rightarrow \mathbb C$. In addition, if $z\in \mathbb C$ then $z= z_R+iz_I$.

We are interested in two wide classes of the complex potentials that can be associated with the eigenvalue problem (\ref{H}--\ref{es}), the first one will be called {\em continuous class} and is defined by the conditions

\begin{enumerate}
\item[i)] 
$V_I(x)$ is a continuous function in $\mathbb R$ that changes sign only once. 
\item[ii)]
If $V_I =0$ is fulfilled in any ${\cal I} \subset \mathbb R$, then ${\cal I}$ is of measure zero.
\end{enumerate}

\noindent
The second class includes complex potentials of the form
\be
V(x) = {\cal V}(x) [ \Theta (x+a)-\Theta (x-b)], \quad a \geq 0, \quad b \geq 0,
\label{potwell}
\ee
where the complex-valued function ${\cal V}: \mathbb R \rightarrow \mathbb C$ satisfies the conditions

\begin{enumerate}
\item[iii)] 
${\cal V}_I(x)$ is allowed to be a piecewise function in $\mathbb R$. This changes sign only once in $(-a,b) \subseteq \mathbb R$. 
\item[iv)]
If ${\cal V}_I =0$ is fulfilled in any ${\cal I}_P \subset (-a,b)$, then ${\cal I}_P$ is of measure zero.
\end{enumerate}

\noindent
The set of these last potentials will be called {\em short-range class}.

\subsection{Main results}

Assuming that the eigenvalue problem (\ref{H}--\ref{es}) has been solved for a given potential of either the continuous or the short-range classes, the following results apply.

\noindent
{\sc Lemma 2.1}  {\em If $\psi(x)$ is a  solution of the eigenvalue problem}~(\ref{H}--\ref{es}) {\em then the Wronskian $W[\psi_R, \psi_I]$ is different from zero for all $x \in \mathbb R$.}

This result establishes the linear independence between $\psi_R$ and $\psi_I$, a fundamental property of the $\psi$--functions in (\ref{H}--\ref{es}). Thus, although $\psi_R$ and $\psi_I$ are associated with the same energy eigenvalue $E_k$, they are not equivalent. In particular, this last means that $\psi_R = c \,\psi_I$, with $c$ an arbitrary (not null) number, is not possible\footnote{Of course, as there is no degeneracy for bound states in one-dimensional real potentials, if $V(x) \in \mathbb R$ then $\psi_R \propto \psi_I$, by necessity.}. As an immediate consequence we find that the solutions $\psi(x)$ of the eigenvalue problem (\ref{H}--\ref{es}) are free of zeros on the real line.

\noindent
{\sc Corollary 2.1}  {\em The zeros of $\psi_R$ and  $\psi_I$ in $\mathbb R$, if they exist, do not coincide.}

The absence of nodes (zeros on the real line) is a common profile of the eigenfunctions of  solvable complex-valued potentials with real spectrum, even if such potentials are ${\cal PT}$-invariant (some examples and references are given in the next sections).

\noindent
{\sc Theorem 2.1} {\em Let $\psi(x)$ be a normalizable solution of the eigenvalue problem}~(\ref{H}--\ref{es}). {\em If $\lambda_1 < \lambda_2 < \cdots$ and $\mu_1 < \mu_2 < \cdots$ are respectively the zeros of $\psi_R$ and $\psi_I$ in $\mathbb R$, then either
\be
\cdots < \lambda_{\ell} < \mu_{\ell} < \lambda_{\ell+1} < \mu_{\ell+1} < \cdots \quad \mbox{or} \quad \cdots < \mu_{\ell} < \lambda_{\ell} < \mu_{\ell+1} < \lambda_{\ell+1} < \cdots.
\ee
That is, the zeros of $\psi_R$ and $\psi_I$ interlace in $\mathbb R$.}

This last is the main result of the present section. As $\psi(x)$ is free of nodes, the probability density $\rho(x)=\vert \psi(x) \vert^2 = \psi_R^2(x) + \psi_I^2(x)$ is different from zero in $\mathbb R$. Besides, the real (imaginary) part of the wave function does not contribute to the probability density at the points $\lambda_{\ell}$ ($\mu_{\ell}$). Therefore, the profile of $\rho(x)$ is mainly determined by the interlacing of $\lambda_{\ell}$ and $\mu_{\ell}$.

Although the above results permit the study of the distribution of  zeros of $\psi_R$ and $\psi_I$ in $\mathbb R$, a priori we can not say anything about the number of such zeros. As we are going to see, depending on the case, they can be finite or infinite. Nevertheless, assuming that the number of zeros of the real and imaginary parts of the wave function is finite, we have the following result.

\noindent
{\sc Corollary 2.2} {\em If $n_R$ and $n_I$ are respectively the number of zeros of $\psi_R$ and  $\psi_I$, then}
\be
\vert n_R - n_I \vert \leq 1.
\ee
That is, if $n_I$ and $n_R$ are finite then they differ by at most one unit.

\subsection{Applications}

In this section we present some examples addressed to show the applicability of the above results. We distinguish between ${\cal PT}$-invariant and non ${\cal PT}$-symmetric potentials. The examples include cases with either finite or infinite number of energy eigenvalues as well as wave functions with real and imaginary parts that have either a finite or an infinite number of zeros.

\subsubsection{${\cal PT}$-symmetric potentials}
\label{potpt1}

The quantum systems that are invariant under the space-time reflection (${\cal PT}$) are called ${\cal PT}$-symmetric \cite{Ben99,Ben05}. The action of the parity reflection ${\cal P}$ on the position and momentum operators is ruled by the transformation ${\cal P}: (x,p) \rightarrow (-x,-p)$. In turn, the time reversal ${\cal T}$ operates also on the imaginary number $i$ as ${\cal T}:  (x,p,i) \rightarrow (x,-p,-i)$. Thereby, given a complex-valued potential $V(x)$, the combined transformation produces ${\cal PT}: V(x) \rightarrow V^*(-x)$, with $z^*$ the complex conjugate of $z \in \mathbb C$. The examples presented in this section are invariant under the space-time reflection, i.e. $V(x)=V^*(-x)$. 

\begin{figure}[htb]
\centering 
\includegraphics[width=0.7\textwidth]{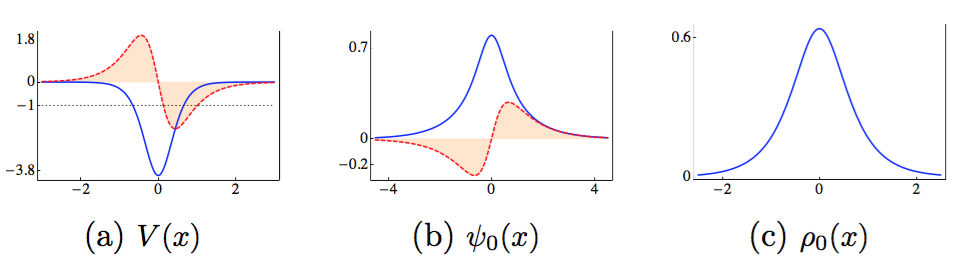}

\caption{\footnotesize 
Real (blue curves) and imaginary (red-dashed curves, filled to the $x$-axis) parts of the P\"oschl-Teller-like potential  (\ref{pot0}) and its single bound state (\ref{func0}) for $\kappa=2$, figures (a) and (b), respectively. The dotted horizontal line in Figure~(a) represents the single discrete energy $E_0=-1$ for this system. The related probability density is free of zeros, as this can be appreciated in Figure (c).}
\label{fig1}
\end{figure}

$\bullet$ As a first example we consider the complex P\"oschl-Teller-like potential shown in Figure~\ref{fig1}(a) and defined by the expression
\be
V(x)= -\left( \frac{\kappa}{\cosh{(\kappa x)}} \right)^2 \left[ 1 + i \sinh{(\kappa x)} \right], \quad \kappa >0.
\label{pot0}
\ee
There is only one square-integrable eigenfunction for this ${\cal PT}$-symmetric potential \cite{Ros15}, this is associated with the energy $E_0= -\tfrac{\kappa^2}{4}$ and given by 
\be
\psi_0(x)= \left( \frac{\kappa}{\pi} \right)^{1/2} \frac{e^{ i\arctan\left[ \tanh{\left(\frac{\kappa x}{2}\right)} \right] }} { \cosh^{1/2} (\kappa x) }.
\label{func0}
\ee

For the sake of clarity, let us verify point by point that the formal results of the previous section are satisfied in this particular case. First notice that $V_I(x)$ is of continuous class and changes sign at $x=0$ only. Now, the Lemma~2.1 is automatically satisfied because the Wronskian
\be
W[\psi_R,\psi_I] =\frac{\kappa^2 }{2\pi}\text{sech}(\kappa x),
\ee
is different from zero for all $x \in \mathbb R$, monotonic increasing in $(-\infty,0)$, and monotonic decreasing in $(0,\infty)$. On the other hand, the zeros of the real and imaginary parts of the eigenfunction (\ref{func0}) are distributed according to the rules
\be
\arctan\left[ \tanh{\left(\frac{\kappa x}{2}\right)} \right] = \pm \left(n + \frac12 \right) \pi, \qquad 
\arctan\left[ \tanh{\left(\frac{\kappa x}{2}\right)} \right] = \pm n\pi,
\label{pt1}
\ee
so that Corollary~2.1 and Theorem~2.1 are true. Moreover, as $\tanh : \mathbb R \rightarrow [-1,1]$, we find that $\psi_R$ has not zeros in $\mathbb R$ while $\psi_I$ has only one at $x=0$, see Figure~\ref{fig1}(b). That is, Corollary~2.2 holds because $n_R=0$ and $n_I=1$.

In the previous section we mentioned that the probability densities of this kind of problems have no zeros in $\mathbb R$. This is very clear in the present case because $\rho_0(x) = \frac{\kappa}{\pi} \mbox{sech} (\kappa x)$ is different from zero for all $x \in \mathbb R$, see Figure~\ref{fig1}(c). Moreover, any particle of energy $E_0=-\tfrac{\kappa^2}{4}$ is localized in the vicinity of the origin (where $V_R(x)$ has a global minimum), in agreement with the distribution of probabilities associated with $V_R(x)$ alone. However, as we are going to see, such agreement is not always true. 

\begin{figure}
\centering
\includegraphics[width=0.5\textwidth]{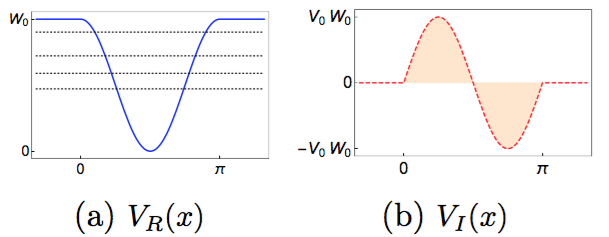}

\caption{\footnotesize Real (a) and imaginary (b) parts of the short-range potential (\ref{pot1}). For 
$W_0=30$ and $V_0=0.49$, this potential admits four bound energies only (dotted horizontal lines in Figure~a). The imaginary part is filled to the $x$-axis.}
\label{fig2A}
\end{figure}

$\bullet$ Our second example is the sinusoidal complex well defined by 
\be
V(x)=\left\{
\begin{array}{cl}
W_0(\cos^2{x}+iV_0\,\sin{2x}), & x\in [0,\pi] \\[2ex]
W_0, & x\notin [0,\pi]
\end{array}
, \quad W_0, V_0 \in \mathbb R.
\right.
\label{pot1}
\ee
The function $V_I(x)$ is of short-range class and changes sign at $x=\pi/2$ only, as it is shown in Figure~\ref{fig2A}(b). In turn, the real part of (\ref{pot1}) is depicted in Figure~\ref{fig2A}(a).

After the change $y=x-ix_0$,  the nontrivial part of the eigenvalue equation (\ref{schro1}) is reduced to the Mathieu equation \cite{Mid10,Sin13}:
\be
\left[ \frac{d^2}{dy^2} + a-2q\,\cos{2y} \right]\psi(y)=0, 
\label{Mat}
\ee
where analytic continuation has been assumed, and
\be
a=E-\frac{1}{2}W_0,\hspace{0.3cm}q=\frac{1}{4}W_0\sqrt{1-4V_0^2},\hspace{0.3cm}x_0=\frac{1}{2}\tanh^{-1}{(2V_0)}.
\ee
If $V_0  \leq \frac12$ then $q \in \mathbb R$, so that the ${\cal PT}$-symmetric potential (\ref{pot1}) has real eigenvalues \cite{Mid10}.  If $V_0>\frac12$, the ${\cal PT}$ symmetry is spontaneously broken and potential (\ref{pot1})  exhibits  anomalous scattering and spectral singularities \cite{Sin13}. We are interested in the former case. 

The general solution is of the form
\be
\psi(x)=\left\{
\begin{array}{cc}
A_3 e^{i\kappa x}+B_3 e^{-i\kappa x}, & x\leq 0\\[2ex]
A_2\,M_C(a,q,x-ix_0)+B_2\,M_S(a,q,x-ix_0), & 0< x< \pi\\[2ex]
A_1 e^{i\kappa x}+B_1e^{-i\kappa x}, & x\geq \pi
\end{array}
\right.
\label{solsin}
\ee
with $M_C(a,q,y)$ and $M_S(a,q,y)$ the Mathieu functions \cite{McL51}, and $\kappa=\sqrt{E-W_0}$. 

As (\ref{pot1}) is a short-range potential, the continuity of $\psi$ and $\psi'$ leads to a relationship between the coefficients $A_3, B_3$ and $A_1, B_1$ of (\ref{solsin}). Symbolically one has
\be
\vert {\cal A}_3 \rangle = M(E) \vert {\cal A}_1 \rangle, \qquad \vert {\cal A}_k \rangle := \left(
\begin{array}{c}
A_k\\ B_k
\end{array}
\right),
\quad k=1,3, \quad \mbox{Det}[M(E)]\neq 0,
\label{transfer}
\ee
where the complex $2\times2$-matrix $M(E)$ is defined by the functions $\psi(x)$ and $\psi'(x)$, evaluated at the boundaries of the interaction zone, and depend on the energy $E$. Assuming that a test particle comes from the right we can take $A_3=0$. It is straightforward to verify that the energy of the bound states is defined by the zeros of the matrix element $M_{11}$ in the interval $0< E <W_0$ (note that $B_1$ is equal to zero for such energies). 

\begin{figure}
\centering
\includegraphics[width=0.7\textwidth]{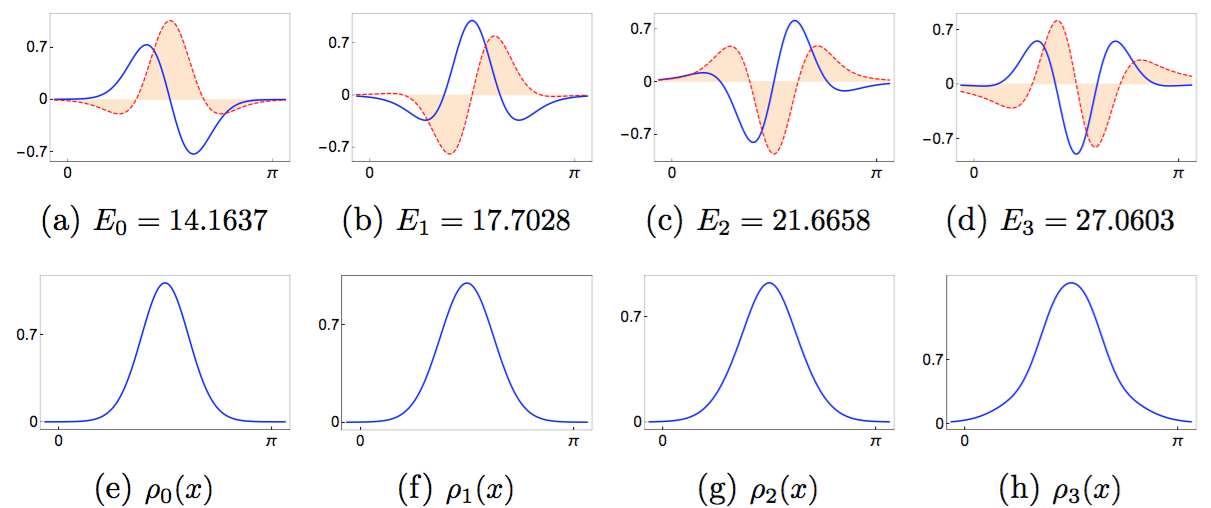}

\caption{\footnotesize The upper row shows the four wave functions of the potential displayed in Figure~\ref{fig2A}. In all cases the imaginary part is in red-dashed (the related curves have been filled to the $x$-axis), and the real part in blue. The lower row shows the corresponding probability densities.}
\label{fig2B}
\end{figure}

For $W_0=30$ and $V_0=0.49$ the potential (\ref{pot1}) has four bound states only. These are shown in Figure~\ref{fig2B}, the corresponding energies were calculated numerically and are indicated in each one of the figure captions. The zeros of $\psi_R$ and $\psi_I$ are reported in Table~\ref{table1}, these have been calculated numerically and ordered by fixing a global phase in each one of the wave functions (the curves depicted in Figure~\ref{fig2B} are consistent with such a phase choice). In this form, Theorem~2.1 and Corollary~2.2 are true.

\begin{table}[htb]
\begin{center}
\scalebox{0.7}{
\begin{tabular}{ |c|l|c|}
\hline
\multicolumn{3}{ |c| }{interlacing of zeros for $\psi(x)$ defined in (\ref{solsin})} \\
\hline
\multirow{2}{1em}{$\psi_0$} & $n_R=1$ & $\lambda_1 = 1.570$ \\ 
& $n_I=2$ & $\mu_1=1.050$,  $\mu_2=2.091$\\ 
\hline
\multirow{2}{1em}{$\psi_1$} & $n_R=2$ & $\lambda_1=1.141$, $\lambda_2=2.000$ \\ 
& $n_I=3$ & $\mu_1=0.511$, $\mu_2=1.571$, $\mu_3=2.631$ \\ 
\hline
\multirow{2}{1em}{$\psi_2$} & $n_R=3$ & $\lambda_1=0.744$, $\lambda_2=1.567$, $\lambda_3=2.388$  \\ 
& $n_I=2$ & $\mu_1=1.196$, $\mu_2=1.938$ \\ 
\hline
\multirow{2}{1em}{$\psi_3$} & $n_R=4$ & $\lambda_1=0.360$, $\lambda_2=1.250$, $\lambda_3=1.894$, $\lambda_4=2.793$ \\ 
& $n_I=3$ & $\mu_1=0.896$, $\mu_2=1.572$, $\mu_3=2.248$ \\ 
\hline
\end{tabular}
}
\caption{\footnotesize  The zeros of the real and imaginary parts of the eigenfunctions depicted in Figure~\ref{fig2B} satisfy Theorem~2.1 and Corollary~2.2.
}
\label{table1}
\end{center}
\end{table}

A remarkable profile of this example is that the probability density of each one of the four bound states is different from zero and has only one maximum. The former property, as indicated above, is a natural consequence of Theorem~2.1. However, it  is unusual to find that a particle in any bound state is always localized in the vicinities of a given point, at least compared with the Hermitian problems for which only the ground state is single peaked. Thus, the absence of nodes derived from Theorem~2.1 produces probabilities which, in general, do not obey the distribution of maxima and minima that is found in the Hermitian problems. This is of particular interest for ${\cal PT}$-symmetric Hamiltonians $H$ because they can be described using a Hamiltonian $\widetilde H$ that is Hermitian with respect to a given positive-definite inner product \cite{Mos03}. In this form, $H$ and $\widetilde H$ are related by a unitary transformation $U_{\rho}^{-1}: {\cal H} \rightarrow \widetilde{\cal H}$, where the new space of states $\widetilde{\cal H}$ is equipped with the same vector space structure as the Hilbert space ${\cal H}$ associated with $H$ (our notation here is slightly different from the one used in\cite{Mos03}). Clearly, $U_{\rho}^{-1}$ and $U_{\rho}$ must operate in such a way that the local properties (e.g., maxima and minima) of the probability density $\rho(x)$ are correctly mapped into the local probabilities of the new density $\widetilde \rho(x)$ and vice versa.

\begin{figure}[htb]
\centering 
\includegraphics[width=.3\textwidth]{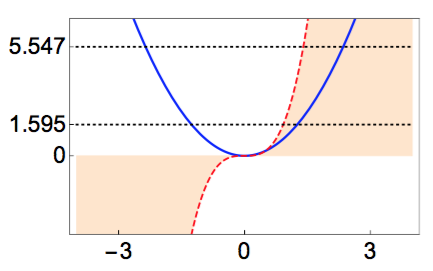} 

\caption{\footnotesize The complex oscillator (\ref{composc}) and its two first energy levels (dotted horizontal lines). The real part is in blue and the imaginary one in red-dashed.
}
\label{fig3A}
\end{figure}

$\bullet$ Another interesting example is given by the ${\cal PT}$-symmetric oscillator shown in Figure~\ref{fig3A} and defined by the expression
\be
V(x)=x^2+ i 2x^3.
\label{composc}
\ee

\begin{figure}[htb]
\centering 
\includegraphics[width=.7\textwidth]{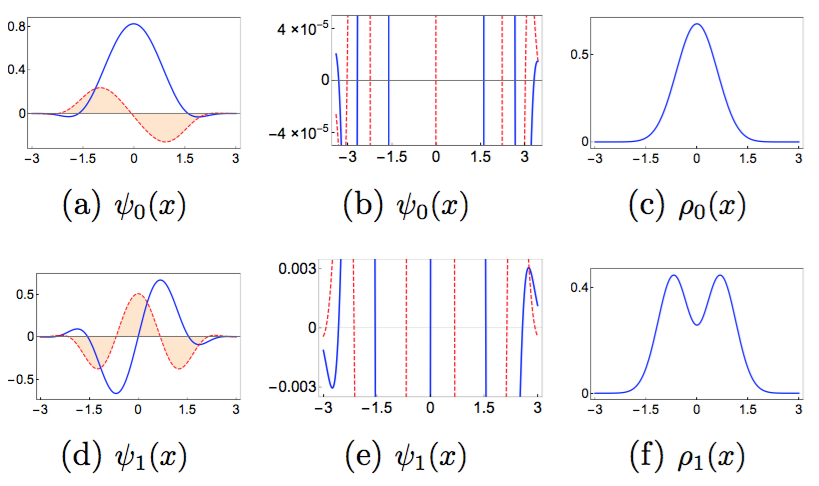} 

\caption{\footnotesize The ground (upper row) and first excited (lower row) states of the oscillator displayed in Figure~\ref{fig3A}. From left to right, the columns exhibit the wave function, the behaviour of the zeros of $\psi_R(x)$ and $\psi_I(x)$, and the probability density. In all cases the real part is in blue and the imaginary one in red-dashed.
}
\label{fig3B}
\end{figure}

\noindent
This last potential is usually interpreted as an oscillator with an imaginary cubic perturbation \cite{Zin10} (see also \cite{Fer14}). After a conjecture by Bessis and Zinn-Justin, it has been proven that the spectrum of (\ref{composc})  is real and positive \cite{Shi02}. Remarkably, both the real and imaginary parts of the corresponding eigenfunctions ``have an infinite number of zeros, individually, when the argument $x$ of the wave function covers the real numbers'' \cite{Nob13}. This can be appreciated in Figure~\ref{fig3B}, where we have depicted the two first bound states. Indeed, all the wave functions exhibit a denumerable set of zeros in their real and imaginary parts, individually, along the real axis. 

The main point here is that the wave functions of potential (\ref{composc}) satisfy Theorem~2.1, even though  $n_R$ and $n_I$ are incommensurable. On the other hand, we would like to emphasize that the probability densities $\rho_n(x)$ of this system behave quite similar to those of the conventional oscillator, with the zeros of the real case substituted by local minima in the complex configuration. Thus, the imaginary part $V_I(x)=2x^3$ affects the global behaviour of a particle by `removing' the points of zero probability associated with $V_R(x)=x^2$, and by displacing the energy eigenvalues $E_n=2n+1$ to the points $E_0= 1.5946$, $E_1= 5.5470,$ etc.

\subsubsection{Non ${\cal PT}$-symmetric potentials}
\label{potnpt1}

In this section we present some examples of complex-valued potentials that have real spectrum but are not invariant under the space-time reflection; that is, $V(x) \neq V^*(-x)$.

\begin{figure}
\centering
\includegraphics[width=.4\textwidth]{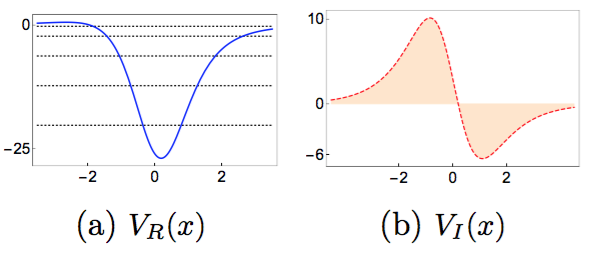} 

\caption{\footnotesize Real (a) and imaginary (b) parts of the non ${\cal PT}$-symmetric potential (\ref{potlev}). For $\nu = -7 +i$, $\mu = -3 -i$, $\epsilon =0.1$ and $\kappa=1$, this potential admits only the five discrete energies represented by dotted horizontal lines in (a). }
\label{fig4A}
\end{figure}

$\bullet$ It can be shown \cite{Lev00} that the eigenfunctions of the potential
\be
V(x)= -\kappa^2\left[ \left( \frac{\nu^2 +\mu^2}{2} -\frac14 \right) \frac{1}{\cosh^2 (\kappa x + i\epsilon)}+i  \left( \frac{\nu^2 -\mu^2}{2} \right) \frac{\sinh{(\kappa x +i\epsilon)}}{\cosh^2{(\kappa x +i\epsilon)}}\right],
\label{potlev}
\ee
are of the form
\be
\psi(x)= c_0 (1-z(x))^{\frac{\nu}{2}+\frac{1}{4}}(1+z(x))^{\frac{\mu}{2}+\frac{1}{4}}P_n^{(\nu,\mu)}(z(x)),
\label{levsol}
\ee
with $P_n^{(\nu,\mu)}$ the Jacobi polynomials \cite{Abr72}, $z(x)=i\sinh{(ax+i\epsilon)}$, and $c_0$ an arbitrary constant. These last functions are regular for $2n + \text{Re}(\nu+\mu)+1 <0$ \cite{Lev00}. That is,  the number of bound states is finite
\be
E_n= - \kappa^2\left(n+\frac{\nu+\mu+1}{2}\right)^2, \qquad n< -\tfrac12 [\text{Re}(\nu+\mu)+1].
\ee
Clearly, for purely imaginary numbers $\nu$ and $\mu$, there are not bound states. If we take $\nu=-7 + i$ and $\mu=-(3 + i)$, the spectrum is real and finite
\be
E_n=-\left(n-\frac{9}{2}\right)^2,\hspace{0.3cm}n=0, 1, 2, 3, 4,
\ee
but potential (\ref{potlev}) is not ${\cal PT}$-invariant \cite{Lev00}, as this can be appreciated in Figure~\ref{fig4A} for $\epsilon =0.1$ and $\kappa=1$. 

\begin{figure}
\centering
\includegraphics[width=.85\textwidth]{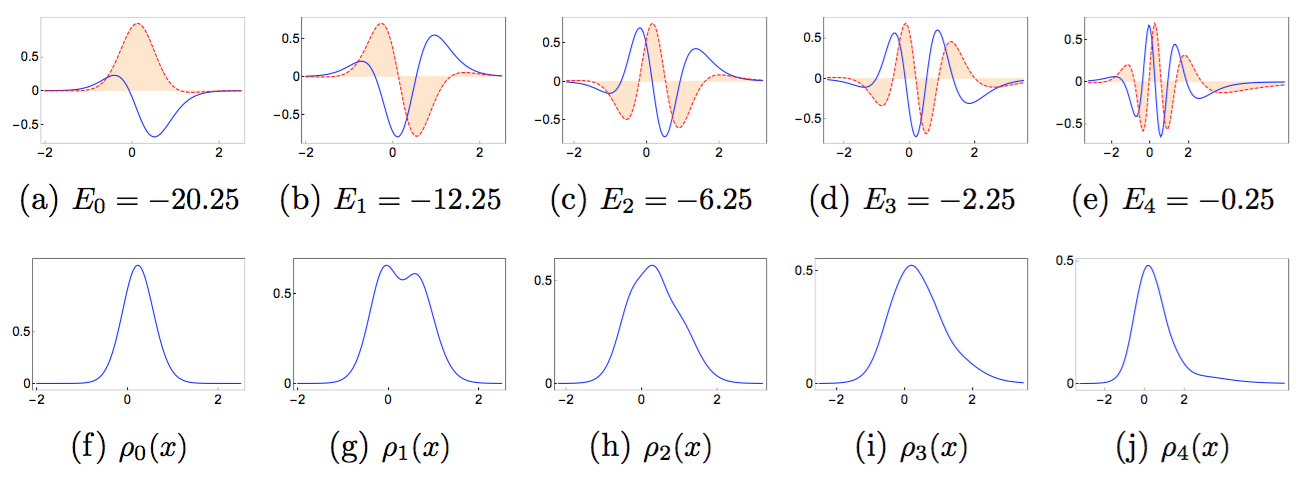} 

\caption{\footnotesize The upper row shows the five eigenfunctions of the potential displayed in Figure~\ref{fig4A}. In all cases the imaginary part is in red-dashed (filled to the $x$-axis) and the real part in blue. The lower row shows the corresponding probability densities.}
\label{fig4B}
\end{figure}

The wave functions and their corresponding energy eigenvalue,  calculated numerically, are displayed in the upper row of Figure~\ref{fig4B}. The zeros of $\psi_R$ and $\psi_I$ have been calculated numerically and ordered by fixing a global phase in each one of the eigenfunctions, they are reported in Table~\ref{table2}. The validity of Theorem~2.1 and Corollary~2.2 is clearly stated.

Note that a particle subjected to the potential displayed in Figure~\ref{fig4A} is localized in the vicinity of the minimum of $V_R(x)$, no matter how excited is its energy, see the lower row of Figure~\ref{fig4B}. The case of the first excited state is slightly different because the probability density $\rho_1(x)$ has two maxima (though they have almost the same value). Thus, the probabilities of this system do not obey the distribution of maxima and minima that is typical in the Hermitian problems.

\begin{table}[htb]
\begin{center}
\scalebox{0.7}{
\begin{tabular}{ |c|l|c|}
\hline
\multicolumn{3}{ |c| }{interlacing of zeros for $\psi(x)$ defined in (\ref{levsol})} \\
\hline
\multirow{2}{1em}{$\psi_0$} & $n_R=1$ & $\lambda_1 = -0.064$ \\ 
& $n_I=2$ & $\mu_1=-1.225$,  $\mu_2=1.110$\\ 
\hline
\multirow{2}{1em}{$\psi_1$} & $n_R=2$ & $\lambda_1=-0.407$, $\lambda_2=0.536$ \\ 
& $n_I=3$ & $\mu_1=-1.372$, $\mu_2=0.127$, $\mu_3=1.327$ \\ 
\hline
\multirow{2}{1em}{$\psi_2$}  & $n_R=3$ & $\lambda_1=-0.700$, $\lambda_2=0.156$, $\lambda_3=0.926$  \\ 
& $n_I=4$ & $\mu_1=-1.577$, $\mu_2=-0.205$, $\mu_3=0.522$, $\mu_4=1.613$ \\ 
\hline
\multirow{2}{1em}{$\psi_3$} & $n_R=4$ & $\lambda_1=-1.009$, $\lambda_2=-0.123$, $\lambda_3=0.533$, $\lambda_4=1.344$ \\ 
& $n_I=5$ & $\mu_1=-1.874$, $\mu_2=-0.496$, $\mu_3=0.204$, $\mu_4=0.896$, $\mu_5=2.016$ \\ 
\hline
\multirow{2}{1em}{$\psi_4$}  & $n_R=5$ & $\lambda_1=-1.394$, $\lambda_2=-0.385$, $\lambda_3=0.264$, $\lambda_4=0.933$, $\lambda_5=1.910$ \\
& $n_I=6$ & $\mu_1=-2.360$, $\mu_2=-0.804$, $\mu_3=-0.048$, $\mu_4=0.581$, $\mu_5=1.352$, $\mu_6=2.698$ \\  
\hline
\end{tabular}
}
\caption{\footnotesize The zeros of the real and imaginary parts of the eigenfunctions depicted in Figure~\ref{fig4B} satisfy Theorem~2.1 and Corollary~2.2.}
\label{table2}
\end{center}
\end{table}

\begin{figure}
\centering
\includegraphics[width=.85\textwidth]{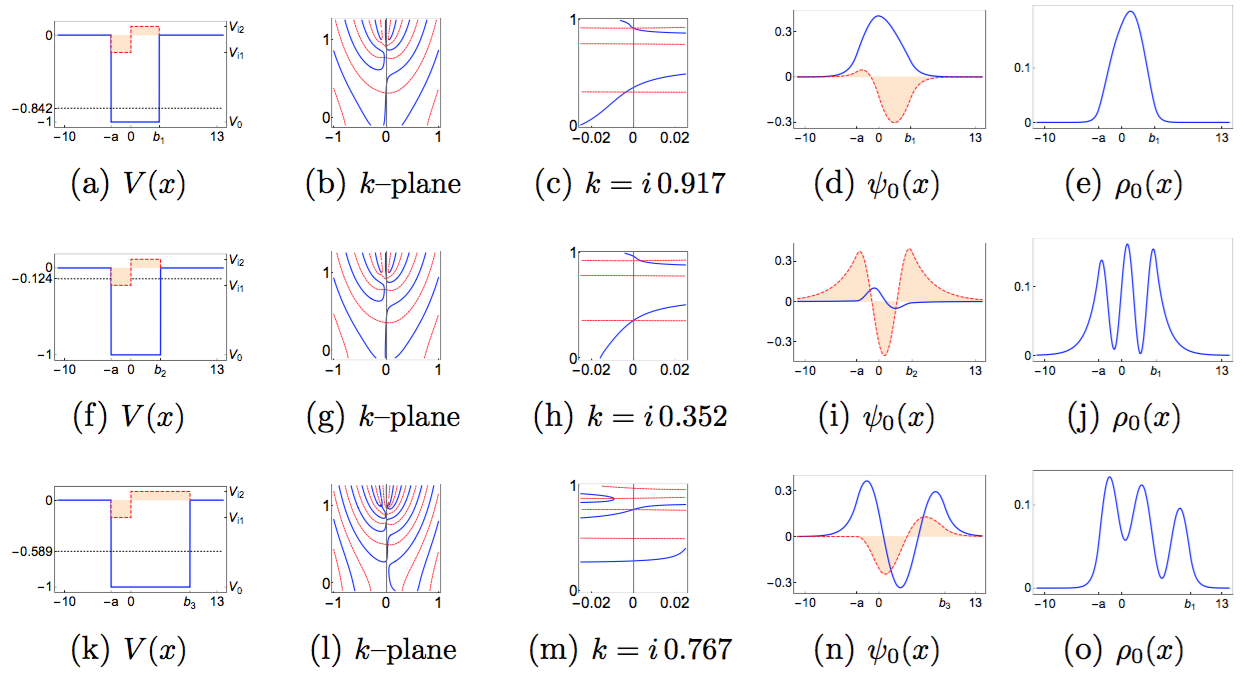} 

\caption{\footnotesize From left to right, the columns of this panel correspond to a complex square-well potential, the complex $k$--plane from which the single bound energy $E=k^2$ is calculated, the point $k=i \kappa$ that defines such bound energy, the eigenfunction, and the corresponding probability density. In all cases the blue and red-dashed curves correspond to the real and imaginary parts respectively. In the first and fourth column the imaginary parts are filled to the $x$-axis. The three rows correspond to three different systems defined by $a=3$, $V_0=-1$, $V_{i1}=-0.2$, $V_{i2}=0.1$, and $b_1=4.2762$ (upper row), $b_2=4.4691$ (middle row) and $b_3=8.9158$ (lower row), respectively. 
}
\label{fig5}
\end{figure}

$\bullet$ An additional example is given by the short-range potential (\ref{potwell}) that is shown in Figure~\ref{fig5}; its real and imaginary parts are respectively given by 
\be
\begin{array}{c}
V_R(x) = V_0 [ \Theta (x-b) -\Theta (x+a)],  \quad V_0>0,\\[2ex]
V_I(x)=  V_{i1} [ \Theta (x) - \Theta(x+a)] + V_{i2} [ \Theta (x) - \Theta(x-b)],  \quad V_{i1} \geq 0, \quad V_{i2} \geq 0.
\end{array}
\label{potwell2}
\ee
For $a=b$ and $V_{i1}= V_{i2} \neq 0$, potential (\ref{potwell2}) is reduced to the ${\cal PT}$-symmetric case reported in e.g. \cite{Zno01}. We are interested in the more general case where $a\neq b$ and  $V_{i1} \neq V_{i2}$ are such that the point spectrum of the complex square well potential (\ref{potwell2}) is real. For simplicity, without loss of generality, we analyze the situation in which there is only one bound state. 
 
Three different examples are shown in the panel of Figure~\ref{fig5}, each one of the rows corresponds to a given potential. Quite interestingly, although these systems have only one discrete energy, the real and imaginary parts of the corresponding eigenfunctions have different number of zeros. Namely, using the parameters $a$, $V_0$, $V_{i1}$ and $V_{i2}$ indicated in Figure~\ref{fig5}, the pairs of numbers $(n_R, n_I)$ are respectively $(0,1)$, $(1,2)$ and $(2,1)$ for the potentials defined by  $b_1= 4.2762$,  $b_2= 4.4691$ and $b_3= 8.9158$, see Table~\ref{table3}. We would like to emphasize that only these values of the parameter $b$, in the interval $b \in (4.267, 8.965)$, give rise to the configuration of a single square-integrable wave function. The wave functions are depicted in the fourth column (from left to right) of Figure~\ref{fig5}; the corresponding energies were calculated numerically by identifying the zeros of $M_{11}$ in the complex $k$-plane, see Eqs.~(\ref{transfer}) as well as the second and third columns of the panel. Clearly, Theorem~2.1 and Corollary~2.2 are true in all these cases.

\begin{table}[htb]
\begin{center}
\scalebox{0.7}{
\begin{tabular}{ |c|l|c|}
\hline
\multicolumn{3}{ |c| }{interlacing of zeros for the single eigenfunction of (\ref{potwell2})} \\
\hline
\multirow{2}{1em}{$b_1$} & $n_R=0$ &  \\ 
& $n_I=1$ & $\mu_1=-1.1086$\\ 
\hline
\multirow{2}{1em}{$b_2$} & $n_R=1$ & $\lambda_1=0.9011$ \\ 
& $n_I=2$ & $\mu_1=-0.9627$, $\mu_2=2.4146$ \\ 
\hline
\multirow{2}{1em}{$b_3$} & $n_R=2$ & $\lambda_1=0.6026$, $\lambda_2=5.3171$  \\ 
& $n_I=1$ & $\mu_1=3.7825$ \\ 
\hline
\end{tabular}
}
\caption{\footnotesize  The zeros of the real and imaginary parts of the eigenfunctions depicted in Figure~\ref{fig5} satisfy Theorem~2.1 and Corollary~2.2.
}
\label{table3}
\end{center}
\end{table}

As we can see, the number of zeros of $\psi_R$ and $\psi_I$ is not directly connected with the number of energy eigenvalues. In contrast, for a real potential and a given discrete energy $E_k$, the oscillation theorem indicates with certainty that $n_R=n_I=k$. Then, for only one discrete energy, the single wave function associated with a particle subjected to a real potential is free of nodes. In our case, the three eigenfunctions displayed in Figure~\ref{fig5} are indeed free of nodes, but this is a general property derived from Theorem~2.1, no matter the number of energy eigenvalues. Moreover, the behaviour of a particle bearing any of these single eigenfunctions depends on the configuration of the potential (\ref{potwell2}).  That is, in the case exhibited in the upper row of Figure~\ref{fig5}, the particle is localized in the vicinity of the origin. This is not the case for the middle and lower rows because the particle can be found, with high probability, around three different points. Even more, it is most probable to find the particle in the neighbourhood of $x=0$ and $x=-a$ for the configuration of the middle and lower rows, respectively.

\section{Susy-generated complex potentials}
\label{darbouxsec}

In this section we address our analysis to the complex-valued potentials that are generated as the Darboux transformation of a given real potential with very well known spectral properties. The main advantage of this approach relies on the fact that some of the oscillation properties of the initial solutions are inherited to both, the imaginary and the real parts of the new solutions.

Let us take a Hermitian Hamiltonian
\be
h= -\frac{d^2}{dx^2} + \vartheta  (x),
\ee
the eigenvalues and square-integrable eigenfunctions of which satisfy
\be
h \varphi_n(x) = {\cal E}_n \varphi_n (x), \quad {\cal E}_0 < {\cal E}_1 < \cdots.
\ee
The Darboux transformation
\be
V(x) = \vartheta (x) + 2 \beta'(x)
\label{d1}
\ee
defines a new eigenvalue problem
\be
H \psi_n (x) = E_n \psi_n (x), \quad E_0 < E_1 < \cdots,
\label{d2}
\ee
with normalized solutions 
\be
\psi_{n+1} (x) = C_{n+1} [\varphi'_n(x) + \beta(x) \varphi_n(x)], \quad E_{n+1} ={\cal E}_n, \quad n=0,1,2,\ldots,
\label{sol}
\ee
whenever 
\be
\psi_0(x) =C_0 \exp \left[-\int^x \beta(y) dy \right]
\label{d3}
\ee
is normalized and $\beta(x)$ satisfies the Riccati equation
\be
-\beta'(x) + \beta^2(x) = \vartheta (x) -E_0.
\label{d4}
\ee
In the above expressions the constants $C_n$ stand for normalization. The new Hamiltonian $H= -\frac{d^2}{dx^2} + V(x)$ is Hermitian, provided that $\beta(x)$ is real and $\beta'(x)$ is free of singularities. The above results are a natural consequence of factorizing the Hamiltonians $h$ and $H$ as the product of two properly defined first order differential operators \cite{Mie84,And84,Fer84}. It is usual to say that $h$ and $H$ are supersymmetric partners because they can be seen as the elements of a matrix Hamiltonian $H_{SS}$ that represents the energy of a system with unbroken supersymmetry \cite{And85,And93,Bag00,Coo01, Mie04}. The state represented by $\psi_0(x)$ is called `missing' because the corresponding energy $E_0$ is absent in the spectrum of the initial Hamiltonian $h$. By construction, such a function is orthogonal to all the other solutions and does not obey the rule (\ref{sol}), so that it could wear properties that are different from those of the states $\psi_{n+1}(x)$. This fact was reported for the first time in \cite{Mie84} and is systematically found in a wide variety of supersymmetric approaches (see, e.g. \cite{Fer84,Mie04,Fer98, Ros98a,Ros98b,Dia99,Mie00,Fer02,Con08}). 

On the other hand, it can be shown \cite{Ros15}  that the superpotential
\be
\beta(x) = -\frac{\alpha'(x)}{\alpha(x)} + i \frac{\lambda}{\alpha^2(x)}, \quad \lambda \in \mathbb R,
\label{super1}
\ee
leads to exactly solvable complex potentials 
\be
V(x) = \vartheta (x) - 2 \frac{d^2}{dx^2} \ln \alpha (x) - i 4 \lambda \, \frac{\alpha'(x)}{\alpha^3(x)}
\label{potD}
\ee
with real spectrum $E_0 <E_1< \cdots$ (other supersymmetric approaches leading to complex-valued potentials with real spectrum can be found in \cite{And99,Bag01,Can98}). Here, $\alpha$ is a real solution of the Ermakov equation
\be
\alpha''(x)= [\vartheta (x) - E_0] \alpha(x)+\frac{\lambda^2}{\alpha^2(x)}
\label{erma}
\ee
that is nonnegative and free of zeros. This can be written as
\be
\alpha(x)=\sqrt{a\,v^2(x)+b\,v(x)z(x)+c\,z^2(x)},
\label{alpha}
\ee
where 
\be
a=\frac{c_0}{w_0^2}, \quad b=2\frac{c_1}{w_0}, \quad c=\frac{\lambda+c_1^2}{c_0}, \quad  b^2-4ac=-4\frac{\lambda}{w_0^2}, \quad w_0 =W[z,v],
\ee
with $c_0$ and $c_1$ arbitrary constants. The functions $z(x)$ and $v(x)$ are two linearly independent solutions of (\ref{erma}) for $\lambda=0$. 

We would like to emphasize that the Darboux transformation defined in (\ref{potD}) works very well for any initial potential $\vartheta (x)$. For instance, the complex P\"oschl-Teller-like potential (\ref{pot0}) is the supersymmetric partner of the free particle potential $\vartheta (x)=0$ for the appropriate parameters. Additional  examples can be consulted in Ref.~\cite{Ros15}. 

Remark that the imaginary part of the potential (\ref{potD}) changes sign as $\alpha'(x)$. In addition, if $x=\xi$ is a zero of $\alpha'(x)$ then
\be
V_R(\xi) = \vartheta(\xi) -2 \frac{\alpha'' (\xi)}{\alpha(\xi)}, \qquad V_I(\xi)=0.
\label{proppot}
\ee
That is, the zeros of $\alpha'(x)$ are zeros of $V_I(x)$ and extremal points of $V_R(x)$. Therefore, depending on the explicit form of $\vartheta(x)$ and $\alpha''(x)$, the points $x=\xi$ determine the maxima and minima of  $V_I(x)$. The fine-tuning of the parameters permits to take $\alpha(x)$ with a slope $\alpha'(x)$ that changes sign only once in $\mathbb R$. Hence, the complex-valued potential (\ref{potD}) can be always constructed such that it is either ${\cal PT}$-invariant or non ${\cal PT}$-symmetric and satisfies the assumptions (i)--(ii) of Section~\ref{Inter}.

\subsection{Main results}

Given a complex-valued potential (\ref{potD}), constructed to satisfy the assumptions (i)--(ii) of Section~\ref{Inter}, the following results apply.

\noindent
{\sc Lemma 3.1} {\em Let $\psi_{n+1} (x)$ be the normalizable eigenfunction of potential} (\ref{potD}) {\em belonging to the eigenvalue $E_{n+1}$. If $n_R$ and $n_I$ denote respectively the number of zeros of $\mbox{\rm Re} (\psi_{n+1})$ and $\mbox{\rm Im} (\psi_{n+1})$ in $\mathbb R$, then}
\be
n_R \geq n+1, \quad n_I = n, \quad n=0,1,2,\ldots
\ee

Thus, there is a lower bound for the zeros of the real and imaginary parts of the $(n+1)$th excited state. The latter has exactly $n$ zeros and the former has no less than $n+1$ zeros. Remark that the missing state $\psi_0(x)$ is excluded from the applicability of Lemma~3.1 due to the peculiarities of its construction that were discussed above.

\noindent
{\sc Theorem 3.1} {\em The normalizable eigenfunction $\psi_{n+1}(x)$ of $H=-\frac{d^2}{dx^2} + V(x)$, with $V(x)$ defined by} (\ref{potD}), {\em has not nodes and the zeros of its real and imaginary parts interlace in $\mathbb R$. In addition, 
\be
n_R = n+1, \quad n_I =n, \quad n=0,1,2,\ldots,
\ee
where $n_R$ and $n_I$ are respectively the number of zeros of $\mbox{\rm Re} (\psi_{n+1})$ and $\mbox{\rm Im} (\psi_{n+1})$ in $\mathbb R$.}

Then, given the energy eigenvalue $E_{n+1}$, the real and imaginary parts of the wave function  (\ref{sol}) have a definite number of zeros in $\mathbb R$. Besides, the distribution of the zeros of $\mbox{\rm Re} (\psi_{n+1})$ is similar to the distribution of the nodes of $\varphi_{n+1}(x)$, while the zeros of  $\mbox{\rm Im} (\psi_{n+1})$ are distributed as the nodes of $\varphi_n(x)$.  As a consequence, the probability density $\rho_{n+1}(x)=\vert \psi_{n+1}(x)\vert^2$ is free of zeros in $\mathbb R$ and has local minima at the points $\lambda_{\ell}$ where $\mbox{\rm Re} (\psi_{n+1})$ is null. This last statement will be clear in the next sections, where we are going to analyze some examples.

\subsection{Applications}

We shall concentrate on the complex supersymmetric partners of the linear harmonic oscillator that have real spectrum. That is, we use $\vartheta (x)=x^2$ in (\ref{potD}) to get
\be
V(x)=x^2-2-2\frac{d}{dx}\left[\frac{2c_1+c_0\,\text{Erf}\,(x)-i\sqrt{\pi}\lambda}{\sqrt{\pi}\,\alpha^2(x)}\right],
\label{miosc}
\ee
where $\mbox{Erf}(x)$ is the error function \cite{Abr72} and 
\be
\alpha(x)=e^{x^2/2}\left[\frac{1}{4}\pi c_0 \text{Erf}\,^2(x)+\sqrt{\pi}c_1\text{Erf}\,(x)+\frac{c_1+\lambda}{c_0}\right]^{1/2}.
\label{aldar}
\ee
The reason for using the complex-valued oscillator (\ref{miosc}) is two-fold: This is general enough to represent all the properties of the Darboux-deformations (\ref{potD}) and, as we have mentioned, the $\alpha$-functions (\ref{aldar}) can be manipulated to get a potential (\ref{miosc}) that is either ${\cal PT}$-invariant or non ${\cal PT}$-symmetric.

\subsubsection{${\cal PT}$-symmetric potentials}
\label{potpt2}

\begin{figure}
\centering
\includegraphics[width=\textwidth]{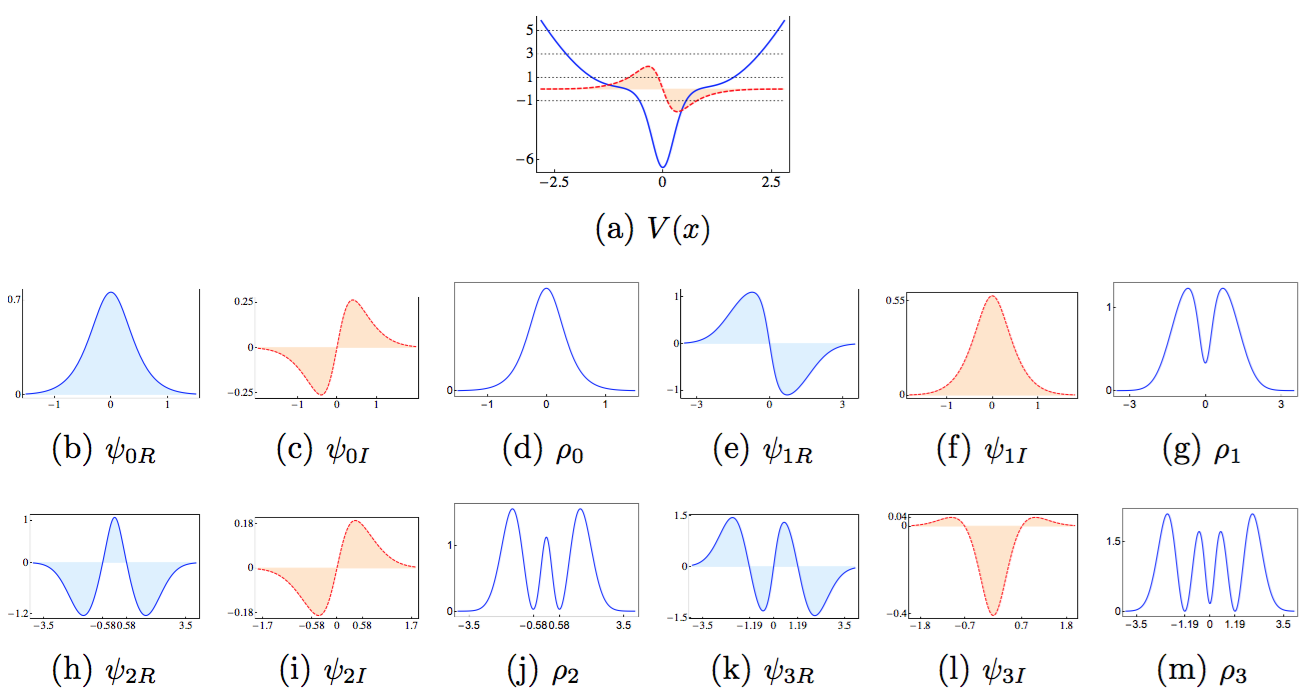} 

\caption{\footnotesize A ${\cal PT}$-invariant version of the complex-valued oscillator (\ref{miosc}) is shown in (a). The panel in the lower rows includes the first four bound states as follows (from left to right): the functions $\psi_R(x)$ are depicted in columns one and four, the functions $\psi_I(x)$ are shown in columns two and five, and the probability densities $\rho(x)$ are exhibited in columns three and six, respectively. In all cases the real part is in blue and the imaginary one in red-dashed. The curves that change sign are filled to the $x$-axis.}
\label{fig9}
\end{figure}

In Figure~\ref{fig9}(a) we show a ${\cal PT}$-invariant version of the complex-valued oscillator (\ref{miosc}), this is obtained for $c_0=2$, $c_1=0$, and $\lambda=1.7$. As indicated above, the spectrum of this potential is equidistant $E_n=2n-1$, with $n=0,1,\ldots$ The wave functions $\psi_{n+1}(x)$ satisfy the Theorem~3.1, this is verified in Table~\ref{table4} were the numerically calculated values of $\lambda_{\ell}$ and $\mu_{\ell}$ are reported. The functions $\psi_{kR}:= \mbox{Re}(\psi_k)$ and $\psi_{kI}:= \mbox{Im}(\psi_k)$ are depicted in the panel of Figure~\ref{fig9} for $k=0,1,2,3$. Notice that the missing state $\psi_0(x)$ does not obey the statement of Theorem~3.1, although the zeros of $\psi_{0R}(x)$ and $\psi_{0I}(x)$ formally interlace because $\lambda_0$ is absent.

On the other hand, the probability densities $\rho_n(x)$, depicted in columns three and six (from left to right) of the panel of Figure~\ref{fig9}, are such that the number of their maxima increases as the level of the energy. That is, $\rho_0(x)$ is localized (single peaked) at origin, $\rho_1(x)$ has two maxima located symmetrically around the origin, and so on. The distribution of these maxima is quite similar to the well known distribution of probabilities in the Hermitian problems. The main difference is that the zeros of probability appearing for the real oscillator have been `removed' and substituted by local minima in the complex-valued oscillator. As this last effect disappears by taking $\lambda=0$, we know with certainty that such a behaviour of $\rho_n(x)$ is due to the imaginary interaction $V_I(x)$. Remember that we have found a similar result for the `complex-perturbed'  oscillator (\ref{composc}). However, there are at least two main differences between these two complex oscillators. The first one is that potential (\ref{miosc}) has the same equidistant spectrum $E_n=2n-1$ for any $\lambda \neq 0$. In turn, the spectrum of the oscillator (\ref{composc}) has to  be evaluated numerically, and it is such that the allowed energies are displaced versions of the eigenvalues $E_n$. Another difference is that, although the imaginary term $2x^3$ of (\ref{composc}) has been considered as a perturbation \cite{Zin10,Fer14}, it is clear that $\vert 2 x^3 \vert$ grows faster than $x^2$ as $\vert x \vert \rightarrow +\infty$. Therefore, the complex oscillator (\ref{composc}) cannot be formally interpreted as  an oscillator with `an imaginary cubic perturbation'. Moreover, it is not clear how to interpret such unbounded term in the potential. In contrast, the imaginary part $V_I(x)$ of our complex oscillator (\ref{miosc}) is bounded, and it is regulated by the parameter $\lambda$ in such a form that $V_I(x)$ can be turned off by making $\lambda=0$. These last properties facilitate the interpretation of $V_I(x)$ as a rightful perturbation (for $\vert \lambda\vert <<1$) which would be associated with dissipation (see e.g. \cite{Cru15,Cru16}).

\begin{table}[htb]
\begin{center}
\scalebox{0.7}{
\begin{tabular}{ |c|l|c|}
\hline
\multicolumn{3}{ |c| }{interlacing of zeros for the wave functions of Figure~\ref{fig9}.} \\
\hline
\multirow{2}{1em}{$\psi_0$} & $n_R=0$  & \\
& $n_I=1$ & $\mu_1=0$ \\
\hline
\multirow{2}{1em}{$\psi_1$} & $n_R=1$ & $\lambda_1=0$ \\
& $n_I=0$ & \\
\hline
\multirow{2}{1em}{$\psi_2$} & $n_R=2$ & $\lambda_1=-0.586$, $\lambda_2=0.586$  \\
& $n_I=1$ & $\mu_1=0$ \\
\hline
\multirow{2}{1em}{$\psi_3$} & $n_R=3$ & $\lambda_1=-1.195$, $\lambda_2=0$,
$\lambda_3=1.195$ \\
& $n_I=2$ & $\mu_1=-0.707$, $\mu_2=0.707$ \\
\hline
\end{tabular}
}
\caption{\footnotesize  The zeros of the real and imaginary parts of the eigenfunctions of the $\mathcal{PT}$-symmetric complex oscillator displayed in Figure~\ref{fig9} satisfy the Theorem~3.1.
}
\label{table4}
\end{center}
\end{table}

\subsubsection{Non ${\cal PT}$-symmetric potentials}
\label{potnpt2}

\begin{figure}
\centering
\includegraphics[width=.9\textwidth]{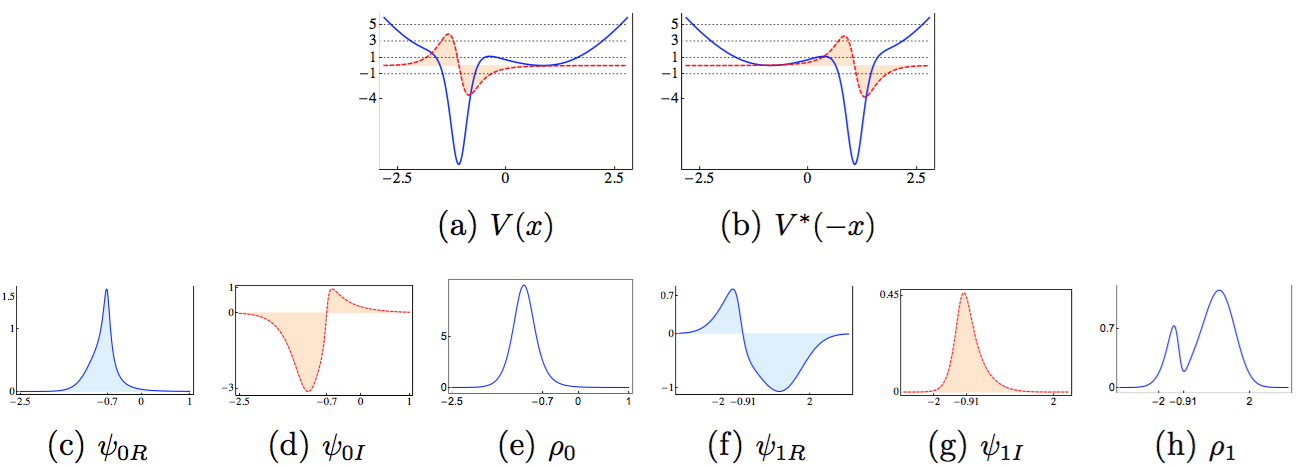} 

\caption{\footnotesize The complex oscillator (\ref{miosc}) with real eigenvalues $E_n=2n-1$ is not ${\cal PT}$-invariant for $c_0=1.2$, $c_1=1$ and $\lambda=0.02$, as this is shown in (a) and (b). The behaviour of the first two bound states can be appreciated in (c)--(e) and (f)--(h) respectively. In all cases the real part is in blue and the imaginary one in red-dashed. The functions $V_I(x)$, $V^*(-x)$, $\psi_R(x)$ and $\psi_I(x)$  are filled to the $x$-axis.
}
\label{fig10}
\end{figure}

\begin{figure}
\centering
\includegraphics[width=.9\textwidth]{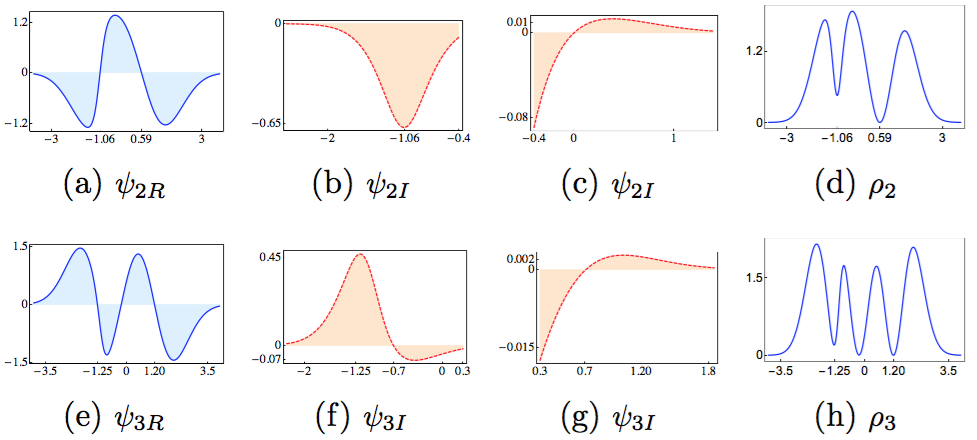} 

\caption{\footnotesize The third and fourth excited states, respectively upper and lower row, of the complex oscillator depicted in Figure~\ref{fig10}. From left to right, the columns include $\psi_R(x)$, a portion of $\psi_I(x)$, the complementary part of $\psi_I(x)$, and the density function $\rho(x)$. In all cases the real part is in blue and the imaginary one in red-dashed. The curves that change sign are filled to the $x$-axis.
}
\label{fig11}
\end{figure}

A non ${\cal PT}$-symmetric version of the complex oscillator (\ref{miosc}) is depicted in Figure~\ref{fig10}(a) for $c_0=1.2$, $c_1=1$, and $\lambda=0.02$. The  ${\cal PT}$-transformed potential $V^*(-x)$ is shown in Figure~\ref{fig10}(b). The zeros of the wave functions belonging to the first four energy levels are reported in Table~\ref{table5}. Clearly, the wave functions $\psi_{n+1}(x)$ satisfy the Theorem~3.1. The functions $\psi_{kR}(x)$ and $\psi_{kI}(x)$ are shown in the lower row of Figure~\ref{fig10} for $k=0,1$, and in the panel of Figure~\ref{fig11} for $k=2,3$.

\begin{table}[htb]
\begin{center}
\scalebox{0.7}{
\begin{tabular}{ |c|l|c|}
\hline
\multicolumn{3}{ |c| }{interlacing of zeros for the wave functions of Figures~\ref{fig10} and \ref{fig11}.} \\
\hline
\multirow{2}{1em}{$\psi_0$} & $n_R=0$  & \\
& $n_I=1$ & $\mu_1=-0.700$ \\
\hline
\multirow{2}{1em}{$\psi_1$} & $n_R=1$ & $\lambda_1=-0.916$ \\
& $n_I=0$ & \\
\hline
\multirow{2}{1em}{$\psi_2$} & $n_R=2$ & $\lambda_1=-1.059$, $\lambda_2=0.590$  \\
& $n_I=1$ & $\mu_1=0$ \\
\hline
\multirow{2}{1em}{$\psi_3$} & $n_R=3$ & $\lambda_1=-1.259$, $\lambda_2=-0.232$,
$\lambda_3=1.201$ \\
& $n_I=2$ & $\mu_1=-0.707$, $\mu_2=0.707$ \\
\hline
\end{tabular}
}
\caption{\footnotesize  The zeros of the real and imaginary parts of the eigenfunctions of the non $\mathcal{PT}$-symmetric complex oscillator displayed in Figure~\ref{fig10}.
}
\label{table5}
\end{center}
\end{table}

The asymmetrical profile of $\psi_{kR}(x)$ and $\psi_{kI}(x)$ is due to the fact that the ${\cal PT}$-symmetry is broken in this case. Thus, the wave functions depicted in Figures~\ref{fig10} and \ref{fig11} can be seen as a deformation of those exhibited in Figure~\ref{fig9}. All the probability densities $\rho_n(x)$ are free of zeros and, as in the previous example, are such that the number of their maxima increases as the level of the energy. Again, the zeros of probability that would be expected in the Hermitian case are removed and substituted by local minima in $\rho_n(x)$.

In this case, the asymmetrical behaviour of the wave functions is preserved after turning off the imaginary interaction. This is because the real part of the potential shown in Figure~\ref{fig10}(a) does not present any symmetry. Indeed, after the translation $x \rightarrow x-\xi$, with $\xi$ the minimum of $V_R(x)$, we see that $V_R(x-\xi)$ is not invariant under the parity reflection ${\cal P}$. In contrast, for the oscillator presented in the previous section, one has $\xi=0$ and $V_R(x) = V_R(-x)$, so that the related functions $\psi_{k+1,R}(x)$ and $\psi_{kI}(x)$ are even or odd according to the value of $k=0,1,\ldots$ In both cases, for $\lambda \rightarrow 0$, the complex-valued oscillator (\ref{miosc}) is reduced to the family of Hermitian oscillators reported by Mielnik \cite{Mie84}, the latter interpreted as a deformation of the harmonic oscillator \cite{Fer94,Fer95,Ros96}. 

As we can see, in general, the Darboux deformed potentials (\ref{potD}) are such that $V_R(x)$ is either invariant or not invariant under parity reflection ${\cal P}$. This last with no dependence on the profile of the imaginary part $V_I(x)$. The symmetrical properties of the corresponding wave functions depend mainly on the parity properties of $V_R(x)$. In any case, the distribution of the maxima of the density probabilities $\rho_n(x)$ is quite similar to that of the Hermitian problems. The points of zero probability that are usual in the Hermitian problems are substituted by local minima of $\rho_n(x)$ in the complex Darboux transformations discussed here.

\section{Proofs}
\label{proofs}

In this section we provide the proofs of the formal results presented in Sections~\ref{Inter} and \ref{darbouxsec}.

{\em Proof of Lemma 2.1.} Decoupling Eq.~(\ref{schro1}) into its real and imaginary parts one obtains 
\be
\psi_R'' + (E-V_R) \psi_R = - V_I \psi_I, \qquad \psi_I'' + (E-V_R) \psi_I = -V_I \psi_R.
\label{system}
\ee
This last system is then reduced to the following differential equation 
\be
W'[\psi_R,  \psi_I]= \vert \psi (x)\vert^2V_I (x).
\label{wrons}
\ee
The Wronskian $W[\psi_R,  \psi_I] := \psi'_R (x) \psi_I (x) -  \psi_R (x) \psi'_I (x)$ is a continuous function of $x$ in $\mathbb R$ because $\psi_R(x)$ and $\psi_I(x)$ are at least $C^1(\mathbb R)$. Hereafter we use the simpler notation $W[\psi_R,  \psi_I] =W(x)$.

In the next steps of the proof we assume that $V(x)$ is of the continuous class.

To investigate the zeros of $W(x)$ in $\mathbb R$ we shall concentrate on its monotonicity properties. With this aim we can assume that the sign of $V_I$ is defined by the rule
\be
sgn(V_I) = \left\{
\begin{array}{rl}
-1, & x< x_0 \\[1ex]
0, &  x=x_0\\[1ex]
1, &  x >x_0\\
\end{array}
\right.
\ee
Therefore, from (\ref{wrons}) we have
\be
sgn(W') = \left\{
\begin{array}{rl}
-1, & x< x_0 \\[1ex]
0, &  x=x_0\\[1ex]
1, &  x >x_0\\
\end{array}
\right.
\ee
That is, the Wronskian $W(x)$ decreases in $(-\infty, x_0)$ and increases in $(x_0, +\infty)$. 

Let us take a pair of points $x_1, x_2 \in (-\infty, x_0)$ such that $x_1 < x_2$. Then either $W(x_1)= W(x_2)$ or $W(x_1)> W(x_2)$. In the former case one has $W(x)=W(x_1)=W(x_2)$, otherwise the Wronskian is not a decreasing function in $(-\infty, x_0)$. Therefore 
\be
W(x_1)= W(x_2) \quad \Rightarrow \quad W'(x)=0 \quad \mbox{for all} \quad x \in (x_1, x_2) \subseteq (-\infty, x_0).
\label{implica}
\ee
However, according to Eq.~(\ref{wrons}), the roots of $W'(x)=0$ are defined by the zeros of either $V_I(x)$ or $\vert \psi(x) \vert^2$. From assumption (ii) we know that $V_I(x)$ cannot be identically zero in $(x_1, x_2)$ because this last subset of $\mathbb R$ is not of measure zero. On the other hand, it can be shown that for nontrivial solutions $\psi(x)$ of the system (\ref{system}), there is not finite interval in $\mathbb R$ for which $\psi_R(x)= \psi_I(x) \equiv 0$ \cite{Inc56}. Thus, the implication (\ref{implica}) is neither associated to the zeros of $V(x)$ nor the zeros of $\psi (x)$. As a consequence, the Wronskian is a monotonic decreasing function in $(-\infty, x_0)$ because $W(x_1)> W(x_2)$ is true for any pair of ordered points $x_1, x_2 \in (-\infty, x_0)$. Using a similar procedure we can show that $W(x)$ is a monotonic increasing function in $(x_0, +\infty)$. 

To determine the sign of $W(x)$ we use the condition (\ref{squareint}) to get
\be
\lim_{\vert x \vert \rightarrow + \infty} W(x) = 0.
\ee
Then, by necessity, in both intervals $(-\infty, x_0)$ and $(x_0, +\infty)$ one has $W(x)<0$. Besides, as $W(x)$ is continuous in $\mathbb R$, it follows that $W(x_0) <0$. Hence, the Wronskian is negative for all $x \in \mathbb R$ and has a minimum at $x_0$. 
\hskip8.5cm $\Box$

\vskip1ex
{\em Proof of Theorem 2.1.} For simplicity, let us suppose that $\psi_R$ has only two zeros $\lambda_1 < \lambda_2$. We will prove by contradiction that there is only one point $\mu$ such that $\psi_I(\mu) =0$ and $\lambda_1 < \mu < \lambda_2$. If $\psi_I$ has no zeros then the function $u=\psi_R \psi_I^{-1}$ is continuous and vanishes at $x=\lambda_1, \lambda_2$. Moreover, $u'(x)$ is continuous in $\mathbb R$ because $\psi_R(x)$ and $\psi_I(x)$ are at least $C^1(\mathbb R)$. Then, by the Rolle Theorem, there exists at least one point $c \in (\lambda_1, \lambda_2)$ such that $u'(c)=0$. However
\be
u'(c) = -\left. \frac{W(x)}{\psi_I^2(x)} \right\vert_{x=c}
\ee
implies that either $\psi_I^2(x)$ diverges at $x=c$ or $W(c)=0$. The former conclusion is not possible as the condition (\ref{squareint}) must be satisfied together with the continuity of $V_I$. In turn, the identity $W(c)=0$ is in contradiction with {\sc Lemma} 2.1. Then $\psi_I$ vanishes at least once in $(\lambda_1, \lambda_2)$. 

Suppose now that $\psi_I$ has at least two zeros in $(\lambda_1, \lambda_2)$. As this last means that $\psi_R$ has no zeros in $(\mu_1, \mu_2)$, the above procedure (with the continuous function $w= \psi_R^{-1} \psi_I$) leads to the conclusion that $\psi_R$ must vanish at least once in $(\mu_1, \mu_2)$, which is a contradiction. Therefore, $\psi_I$ has one and only one zero between two consecutive zeros of $\psi_R$, and vice versa.
\hskip14.5cm $\Box$

\vskip1ex
{\em Proof of Lemma 3.1.} By the Sturm-Liouville theory we know that the wave functions $\varphi_n(x)$ of the real-valued potential $\vartheta(x)$ are complete. Besides, $\varphi_n(x)$ has exactly $n$ zeros in $\mathbb R$. Then, from (\ref{sol}) we see that $\mbox{Re}(\psi_{n+1}) =C_{n+1} \varphi_n'(x)$ and $\mbox{Im}(\psi_{n+1}) =C_{n+1} \beta(x) \varphi_n(x)$. As $\beta(x)$ is free of zeros in $\mathbb R$ the lemma has been proved. 
\hskip5.5cm $\Box$

\vskip1ex
{\em Proof of Theorem 3.1.} It follows from Corollary~2.2 and Lemma~3.1.
\hskip3.0cm $\Box$

\section{Concluding remarks}
\label{concluye}

For complex-valued potentials with real point spectrum, we have found some general properties of the corresponding wave functions $\psi(x)$. The main results have been presented in a series of lemmas, corollaries and theorems that pay attention to the zeros of the real $\psi_R(x)$ and imaginary $\psi_I(x)$ parts of such functions in $\mathbb R$. In particular, it has been proved that the zeros of $\psi_R(x)$ and $\psi_I(x)$ interlace in $\mathbb R$, so that they do not coincide and $\psi(x)$ has not nodes. This last is the reason for which the probability density $\rho(x)=\vert \psi(x) \vert^2$ is free of zeros in $\mathbb R$. 

Although the above properties are present in the wave functions of a wide diversity of complex potentials with real spectrum, we have concentrated on the potentials that satisfy the assumptions (i)--(iv) introduced in Section~\ref{Inter}. Such a classification includes potentials that are ${\cal PT}$-symmetric as well as potentials that are not ${\cal PT}$ invariant. In this context, the examples presented in the previous sections show that the invariance under the ${\cal PT}$ transformation is not a necessary condition to get complex-valued potentials with real spectrum. 

We have found that the symmetrical properties of the imaginary part $V_I(x)$ of the potential have a strong influence on the number and the distribution of the zeros of $\psi_R(x)$ and $\psi_I(x)$. Extreme examples are given by the ${\cal PT}$-symmetric oscillator (\ref{composc}) and the non ${\cal PT}$-symmetric complex square-well potential (\ref{potwell2}). In the former case, discussed in Section~\ref{potpt1}, the imaginary part of the potential $V_I(x)=2x^3$ induces the presence of an infinite number of zeros in $\psi_R(x)$ and $\psi_I(x)$. For an square-well (\ref{potwell2}) with only one bound state, see Section~\ref{potnpt1}, we have shown that the zeros of the real and imaginary parts of the single wave function can be regulated by adjusting the involved parameters. The conclusion is that arbitrary constructions of complex-valued potentials that have real spectrum lead to wave functions with uncontrollable number of zeros in their real and imaginary parts.

To have control on the number and the distribution of zeros of $\psi_R(x)$ and $\psi_I(x)$, it is appropriate to construct complex Darboux deformations of a given real potential $\vartheta(x)$ with well known spectral properties, see Section~\ref{darbouxsec}. In this case, we have shown that the corresponding probability densities $\rho_n(x)$ are such that (1) the number of their maxima increases as the level of the energy (2) the distribution of such maxima is quite similar to that of the Hermitian problems, and (3) the points of zero probability that are usual in the Hermitian problems are substituted by local minima.

The common profile in the complex-valued potentials constructed by a Darboux transformation (\ref{potD}) is that the imaginary part $V_I(x)$ exhibits some symmetries that are controllable by the proper selection of the involved parameters. In particular, we have found that the potentials fulfilling the Theorem~3.1 are such that 
\be
\int_{\mathbb R} V_I(x) dx=0.
\label{area}
\ee
As $V_I(x)$ changes sign only once in $\mathbb R$, this last equation means that an odd function\footnote{Depending of the case, a translation $x \rightarrow x-\xi$ would be necessary if $V_I(x)$ changes sign at $x=\xi \neq 0$. See for instance the discussion in Section~\ref{potnpt2}.} $V_I(x)$ is useful to satisfy the Theorem~3.1 (though $V_I(x)$ is not restricted to be odd). The same property, which we call the {\em condition of zero total area}, is presented in the P\"oschl-Teller-like potential (\ref{pot0}), the complex oscillator (\ref{composc}), and in the ${\cal PT}$-symmetric version of the complex square-well potential (\ref{potwell2}) discussed in the previous sections. All these potentials are such that the probability densities behave as indicated above for the Darboux-deformed potentials. In turn, the other potentials discussed along this work are such that the related probability densities do not obey the distribution of maxima and minima that is usual in the Hermitian problems. This is because either the condition of zero total area (\ref{area}) is not satisfied or the function $V_I(x)$ is not continuous in $\mathbb R$. For example, the potential (\ref{pot1}) satisfies the condition (\ref{area}) but it is of short-range class. As a consequence, a particle in any of the corresponding bound states is localized around $x=\pi/2$, see Figure~\ref{fig2B}. Thus, it seems that both, the condition of zero total area (\ref{area}) and the continuity of $V_I(x)$ in $\mathbb R$ are necessary to ensure a `regular' distribution of maxima and minima in the related probability densities.

Considering the above remarks we have an additional result:

\noindent
{\sc Conjecture} 5.1 {\em Let $V(x)$ be a complex-valued potential with real point spectrum of the continuous class. If the condition of zero total area} (\ref{area}) {\em is satisfied, then the involved probability densities are such that (1) the number of their maxima increases as the level of the energy (2) the distribution of such maxima is quite similar to that of the Hermitian problems, and (3) their local minima correspond to points of zero probability in the Hermitian problems.}

It is clear that the complex potentials that are ${\cal PT}$-invariant satisfy the condition of zero total area in their respective domains of definition. In addition, we have shown that this is a common property in all the complex  Darboux deformed potentials (\ref{potD}). The applicability of Conjecture~5.1 for other complex-valued potentials with real spectrum is open, further insights will be reported elsewhere.

\section*{Acknowledgment}

AJN acknowledges the support of CONACyT (PhD scholarship number 243357)


\end{document}